# Regenerative Soot-V: Spectroscopy of the regenerative sooting discharges

Shoaib Ahmad[1,*]

[1]*National Centre for Physics, Quaid-i-Azam University Campus, Shahdara Valley, Islamabad, 44000, Pakistan*

[*]Email: sahmad.ncp@gmail.com

**Abstract.** The mechanisms and processes of the formation of the regenerative soot in a graphite hollow cathode discharge that produces and emits carbon clusters are presented. Mass spectrometry with a designed $E \times B$ velocity filter analyzes the entire range of the charged clusters from $C_1$ to $\sim C_{4300}$. The state of the carbon vapor within the source is evaluated by using the characteristic line emissions from the carbonaceous discharge whose formative mechanisms depend upon the kinetic and potential sputtering of the sooted cathode. The carbonaceous discharge generates atomic and ionic C and its clusters $C_m$ ($m \geq 2$), noble gas metastable atoms and ions, energetic electrons and photons in the cavity of the graphite hollow cathode. The parameters of soot formation and its recycling depend critically on the discharge parameters, the geometry of the hollow cathode and 3D profile of the cusp magnetic field contours.

PACS. 34.50.Dy Interactions of atoms and molecules with surfaces; photon and electron emission; neutralization of ions – 34.80.Dp Atomic excitation and ionization by electron impact –36.40.Wa Charged clusters

## 1. Introduction

In this communication. we summarize the results of an ex- tended study that has been conducted on the mechanisms of production of the regenerative soot. Carbon clusters are be formed in the carbon vapor that is created in a cusp field, graphite hollow cathode discharge. This carbonaceous discharge is characterized and diagnosed using mass spectrometry of the carbon clusters $C_m^+(m \geq 2)$ emitted from the source. The state of the carbon vapour within the source is monitored by photoemission spectroscopy at various stages of soot formation and its recycling. The regenerative soot has been seen to provide an ideal clustering environment conducive to the formation of all sorts of clusters $C_m^+ (m \leq 10^4)$ including the linear chains, rings, closed cages and even onions [1]. The mass spectrometry has identified the existence of sooting layers on graphite cathode's inner walls as a pre-condition for the formation of clusters. In another study [2] the photon emission spectra of the atoms, molecules and ions (both positive as well as negative) were obtained from the energetic heavy ion irradiation of the non-regenerative, flat graphite surfaces. The velocity spectra of these sputtered clusters $C_1^- \ldots C_4^-$ showed similarities with results obtained by other workers from sublimation [3] or sputtering by ion bombardment [4] of graphite. However, there are significant differences in the formation of sputtered clusters and those that are produced in a regenerative sooting environment. The regenerative soot formation is identified by the state of excitation and ionisation of the carbonaceous discharge. Our indicators are the relative number densities of the excited and ionised neon and carbon. Carbon cluster formation in sooting environments has led to the discovery of fullerenes in the laser ablated graphite plumes [5]. The technique for mass production of $C_{60}$ utilizes high pressure arc discharges between graphite electrodes [6]. This again is a non-regenerative arc discharge and efficient production of the "buckyballs" depends upon the discharge parameters and the soot collection methodology. Whereas, in a graphite hollow cathode discharge one recycles the cathode deposited clusters. Therefore, it is a technique of carbon cluster production that relies on the regeneration of the cathode-deposited soot. By monitoring various stages of this regenerative sequences, we have been able to build a picture of the carbon cluster formation mechanisms. We will present the respective roles played by the neutral, excited and ionic states of $C_1$, $C_2$ and $C_3$ and higher clusters in the formative and sooting stages of the regenerative, carbonaceous discharges.

We utilize a compact $E \times B$ velocity filter for the mass spectrometry of the carbon clusters [7]. On the basis of the analyses of these spectra, we identify the transition from a pure sputtering mode to a sooting one. We will also characterize the state of the carbon vapour during this transformation by using emission spectroscopy of the ex- cited and ionized plasma species. The excited and ionized component of the discharge is crucial for the sustenance of the discharge. In addition, we look at the role of the excited species in the recycling and regeneration of the cathode deposited clusters. We explore the roles played by the discharge parameters like the discharge voltage $V_{dis}$ discharge current $i_{dis}$ and the support gas pressure $P_g$. Kratschmer *et al.* [6] had also identified the gas pressure as a critical parameter for sooting in the arc discharge between graphite electrodes. We evaluate the nature of the carbon vapor and the sources of the regeneration of soot by using different modes of excitation and ionization of the atomic and ionic species by electrons, the estimation of the excitation temperatures $T_{exc}$ for various discharge species and the conspicuous transition from the $C_3$ dominated discharge to a sooting plasma.

We have observed unique cluster forming properties in the regenerative sooting discharge of a graphite hollow cathode in cusp magnetic field contours. In the initial stages of the evolution of the carbonaceous discharge, kinetic sputtering is the main contributor. The second mode of discharge can be classified as the sooting mode which may be associated with high pressure discharges where the density of the ionized species $C^+$ and $Ne^+$ considerably reduces. A constant but gentle surface erosion by potential sputtering dominates this mode. Kinetic sputtering is also taking place and the process of cathode sputtering involves the two mechanisms together. Neutral ($C^*$) and ionised carbon ($C^+$) are the integral constituents of all sooting processes and their densities are indicators of the soot formation on the cathode walls. We envisage the sooting mode to imply a loose agglomeration of carbon clusters on the cathode surface being recycled or regenerated by kinetic sputtering with energies up to 500 eV as well as the collisions of metastable $Ne^*$ atoms with energies $E \sim 16.7$ eV. We believe that the transition from the sputtering proficient regime dominated by $C_3$ to the regenerative soot has been identified and presented in this communication. Furthermore, this information provides us an understanding the mechanisms that are responsible for the formation of clusters including the fullerenes in the regenerative sooting discharges.

## 2. Mechanisms of soot formation

During 1984–1985 carbon soot was redefined after the experiments on the laser ablation of graphite followed by the supersonic expansion of the carbonaceous vapor [5, 8]. The agglomerates of pure carbon clusters so formed include the closed caged carbon clusters-fullerenes. Together these clusters produce what is now universally regarded as the *soot*. Various other techniques have been developed in the last decade for the production of soot. These include high pressure arc discharge by Krätschmer *et al.* [6] as a specialist technique exclusively for the production of $C_{60}$ and $C_{70}$. Electron microscopy of the soot bombarded by high energy electrons has shown that the shelled or carbon onion structures can also be produced [9–11].

Energetic ion irradiation in polymers can induce clustering of carbon atoms that has been observed to lead to optical blackening, electrical conductivity changes and has also been studied for ion induced chemical effects [12]. These authors invoked mechanisms of nuclear as well as electronic energy transfer from ion to the carbon atoms in the solid matrix to explain the ion-induced clustering processes. Orders of magnitude estimate for the size of graphitic islands or carbon rich zones range from 100－500 Å. Similar experiments with MeV heavy ion sputtering of polymers at Uppsala [13] identified the formation of fullerenes in MeV iodine ion bombarded PVDF targets. The fullerene yield measurements as a function of ion fluence indicated clustering to be dependent on ion-induced chemical changes in the polymer. Chadderton *et al.* [14] have reported the synthesis of fullerenes after 130 MeV/amu $Dy^{22+}$ ion bombardment of graphite. Chromatography of their irradiated samples has shown traces of $C_{60}$. At PINSTECH, by using 100 keV $Ar^+$, $Kr^+$ and $Xe^+$ beams on amorphous graphite, we have seen clear evidence of ion induced cluster formation in the energy measurements of the direct recoiling clusters from ion bombarded amorphous graphite surfaces [15]. We varied ion energy and the dose with irradiations done at grazing incidence angles.

## 3. Soot regeneration

Understanding of the mechanisms of clustering may lie in the synthesis of common features of the widely different physical methods of producing soot. The aim of the design of a tunable, soot regenerative source is to create a recyclable carbon vapor environment and to study the formative as well as dissociative stages of a carbon cluster. Such a source has been designed and the technical details have been presented elsewhere [1]. The schematic diagram of the regenerative sooting source is shown in Figure 1. Its distinctive features depend upon the sputtering efficiency of the cathode and the subsequent soot formation properties leading to the clustering of carbon atoms and ions. A steady stream of carbon atoms is sputtered into the glow discharge plasma from graphite hollow cathode surface. The key to the ignition and sustenance of the discharge at neon pressures $\approx 10^{-1}-10^{2}$ mbar is a set of six bar magnets wrapped around the hollow cathode providing an axially extended set of cusp magnetic field contours. Xe, Ar, Ne and He have been used to provide the initial noble gas discharge which transforms into a carbonaceous one as a function of the discharge conditions. The sooting discharge so produced demonstrates a temporal growth in the densities of sputtered carbon atoms and ions as a function of the discharge voltage $V_{dis}$ and current $i_{dis}$ and support gas pressure $P_g$. The ions anchor onto a set of field contours, the direction of their consequent gyratio and clustering probability is determined by collision with electrons, neutral and excited C and the support gas atoms.

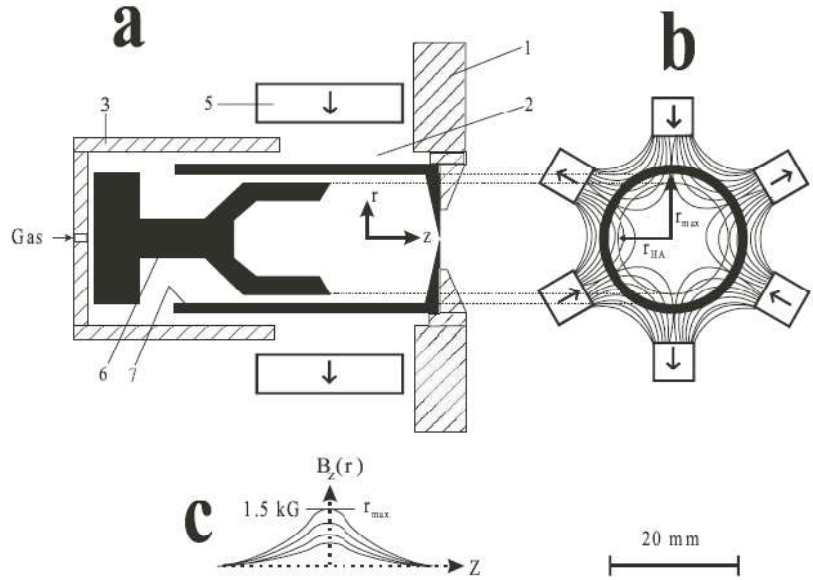

Fig. 1. The schemetic diagram of the source is shown with graphite Hollow Cathode and Hollow Anode. The cusp magnetic field $B_z(r,\theta)$ is also shown with arrows. The numbered items are described in detail in reference [1].

The hexapole field confinement is designed so that the radial $B_r$ and axial $B_z$ field lines produce the combined 3D magnetic field contours $B_z(r, \theta)$ shown in Figure 1c. The streams of gyrating $C^+$, $Ne^+$ and $C^+$ ions with large collision cross-sections eventually lead to the inside walls of the graphite cathode surface where they impact with $E \approx qV_{dis}$, where $q$ is the charge on the ion. The ion-impact continuously modifies the graphite cathode surface properties and it is covered with soot.

These sooted layer carbon cluster emission on the state of sooting in the source. Spectrometry of the regenerative soot is done by a compact, permanent magnet based $E \times B$ velocity filter that has specially been developed for the detection and diagnostics of large carbon clusters [7]. Different mass analyzing techniques can be employed for the detection of such a large range of cluster masses. These include time-of-flight (TOF), momentum analysis and $E \times B$ velocity filtration. For the on-line mass analysis of clusters velocity analysis has certain advantages over the momentum analysis. The TOF technique uses pulsed beams and has superior resolution especially in the higher mass range. However, the $E \times B$ velocity filter has demonstrated its utility as a low cost, useful diagnostic tool for the mass detection of very small to very large clusters as has been demonstrated [1,7].

# 4 Spectroscopy of the regenerative soot

## 4.1 Mass spectrometry with E × B velocity filter

The permanent magnet based $E \times B$ velocity filter can perform mass analysis in a characteristic way. All masses are deflected by the fixed magnetic field according to their respective masses. The straight through beam contains the desired mass at the compensating electric field $\varepsilon_0 = B_0/v_0$, where $B_0$ is the magnetic field intensity along the axis and $v_0$ – the velocity of a particular ion. All charged species including monatomic ions or the ionized clusters are expected to have the same energy. A velocity spectrum always contains all masses irrespective of their mass: the resolvability on the other hand is dependent upon certain design features. Figure 2 shows the experimental arrangement for the detection of carbon clusters $C_m$ from the regenerative sooting source. A well collimated set of extraction lens set up provides a $\pm 0.1°$ beam to the velocity filter of effective length $a$. A picoampere meter measures the analyzed masses on a Faraday cage $l$ mm away from the exit of the filter, in our case $l = 1\,500$ mm. The detection of heavy carbon clusters with an $E \times B$ velocity filter depends on the highest possible magnetic field $B_0$ and that sets other parameters accordingly.

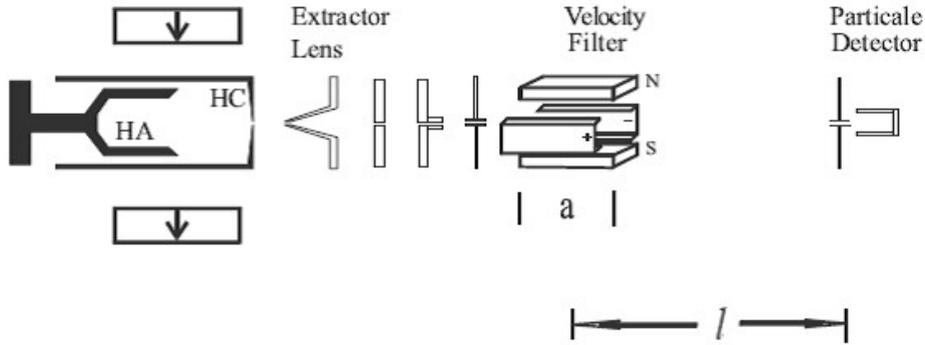

**Fig. 2.** It shows the cluster ion source of Figure 1, extraction lens, collimators, the velocity filter of dimension $a$. The Faraday cage is at distance $l$ from the filter. The source is composed of a graphite Hollow Cathode (HC), Hollow Anode (HA) and a set of hexapole bar magnets shown with arrows.

We have $B_0 = 0.35$ T on the axis of the filter between the poles that are 10 mm apart. Specially shaped electrodes provide the compensating electric field $\varepsilon_0$ for the straight through masses. These electrodes are slightly extended outwards to compensate for the magnet's edge effects. The resolution of the $E \times B$ velocity filter is determined by the dispersion $d$ of masses $m_0 \pm \delta m_0$ from the resolved mass $m_0$ that travels straight through the filter with velocity $v_0 \,(= B_0/\varepsilon_0)$. Dispersion $d \propto al\,(\delta m_0/m_0)(\varepsilon_0/V_{ext})$ where $a$ and $l$ are the lengths of the velocity filter unit and the flight path, respectively. For a given ratio $\delta m_0/m_0$, the dispersion $d$ can be enhanced by stacking multiple filter units since $d \propto na$, $n$ being the number of filter units or increasing the flight path $l$ and also by enhancing $V_{ext}$.

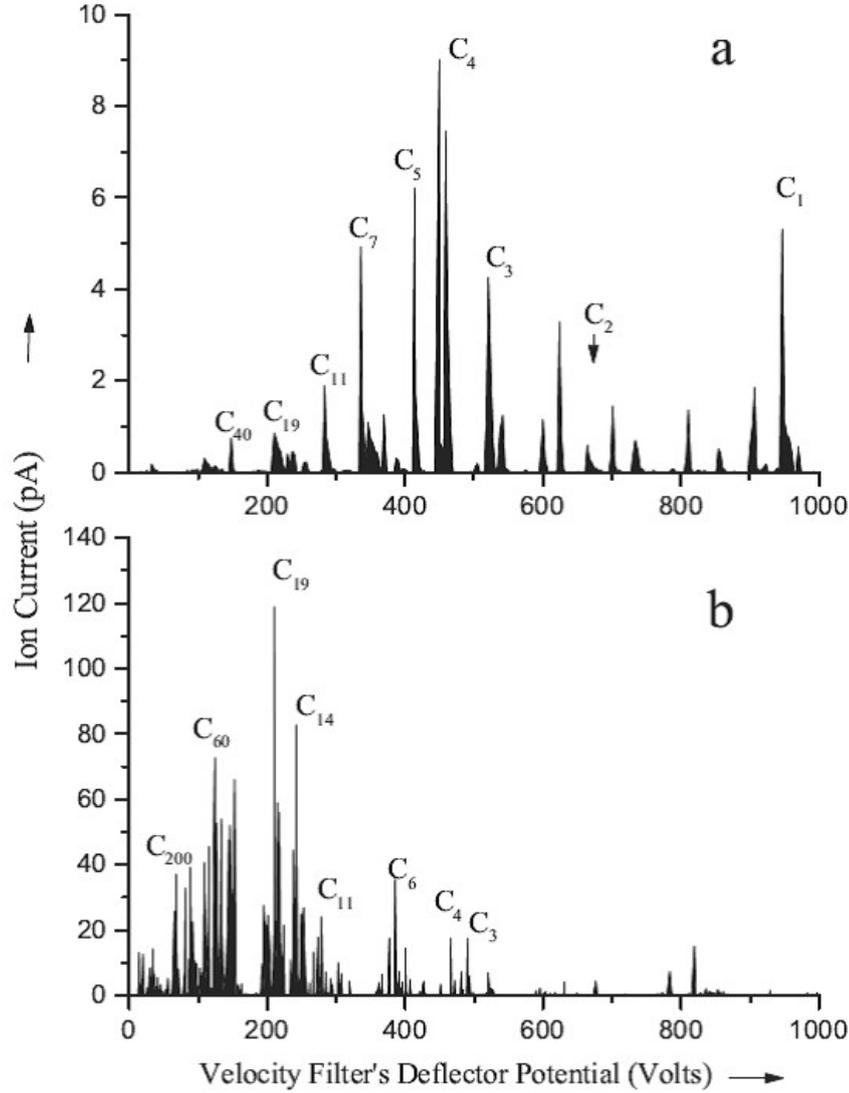

**Fig. 3.** The velocity spectrum of the initial sooting stages of operation with Ne is shown in (a) with pressure in the source $P_g = 2-3\times 10^{-3}$ mbar, $V_d = 0.5$ kV, $I_d = 50$mA and $V_{extraction} = 2$ kV. Clusters from $C_1$ to $C_{19}$ are present in the spectrum. (b) Shows results from a well sooted source operated with Ne at $\approx 60$ watts for 20 hours. It has been obtained with Xe as a source gas. The emitted charged clusters have a range of fullerenes with $m \approx 200$ to $C_{36}$ as well as the rings, chains and linear regimes of clusters. Note the difference in cluster ion intensities in the two spectra.

We have reported [17] the observation of the two distinct stages of sooting; one that is the initial sputtering dominated stage where lower cluster densities of pre- dominantly the linear chains and rings are obtained from $C_1$ to $C_{19}$ as shown in Figure 3a. The later stage of a well sooted discharge is obtained with prolonged operation with high power inputs. It produces the entire range of clusters from linear chains and rings to closed caged fullerenes and perhaps, even the carbon onions. Figure 3b shows the masses higher than $C_{10}$ up to $C_{4350}$. In this spectrum peaks due to $C_{76}$, $C_{60}$, $C_{50}$, $C_{44}$ and $C_{40}$ have a mass difference of $C_2$. Up to $C_{40}$ we have the usual fullerene spectrum. The ring type cluster series starting from $C_{24}$, $C_{21}$, $C_{19}$, $C_{15}$ to $C_{11}$. This is a familiar pat- tern of carbon cluster fragmentation for the fullerenes $C_m$ ($m \geq 32$) and rings and linear chains $C_m$ ($m \geq 24$) and has been seen in the time-of-flight spectrometry of laser ablated graphite experiments [5,6].

## 4.2 Photon emission spectroscopy

Photon emission spectroscopy of the characteristic atomic and molecular lines and bands was done with a compact Jobin Yvon monochromator with a grating blazed at 300 nm and minimum resolution of 1 Å. The quantum efficiency of the photomultiplier tube and the relative efficiency of the grating vary between 180–650 nm. Fused silica window was fitted on the hollow cathode source for the transmission of wavelengths down to ~180 nm. In Figure 4 a typical emission spectrum is presented with the line intensities as obtained from the photomultiplier. But we have multiplied with the appropriate correction factors for the respective wavelengths while calculating the relative number densities of the excited levels. In Figure 4 the graphite hollow cathode discharge with Ne clearly shows three distinct groups of emission lines between 180 and 650 nm. The first group is between 180–250 nm and it includes emission lines belonging to the neutral, singly and doubly charged C. The CI $\lambda = 1931$ Å and $\lambda = 2478$ Å are the signature lines emanating from the same singlet level $^1P_1$. The presence of these lines implies that the initially pure Ne discharge has been transformed into a carbonaceous one. Between 300−400 nm, neon's ionic lines are grouped together with the molecular bands at 357 nm and 387 nm. A significant exception is a NeI line at $\lambda = 3520$ Å which is a resonant line of NeI and we use it along with $\lambda = 5854$ Å for the determination of the excitation temperature $T_{esc}$ of the discharge. The third distinct and high intensity group of emission lines due to the excited atomic NeI lies between 580−650 nm. A large percentage of the discharge power is concentrated in these excited atoms of Ne that cannot de-excite by photoemission to ground. These excited atoms have to give up their energy ~16.7 eV in collisions with the discharge constituents and the sooted cathode walls. We have recently explored their soot regenerative properties as a potential sputtering agent [16]. The emission lines and levels have been interpreted by using NIST's extensive Atomic Spectra Database (ADS) available on the web [18].

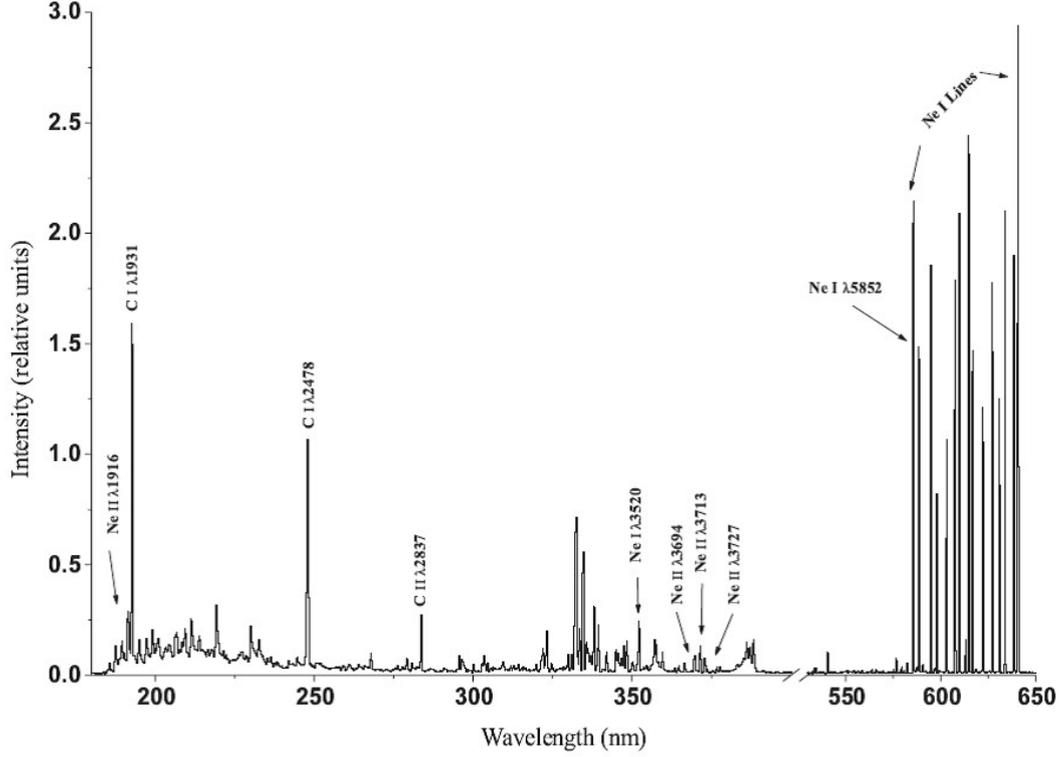

**Fig. 4.** A typical Ne discharge spectrum of the emission lines at $i_{dis}$ = 200 mA. The spectrum shows familiar atomic lines due to CI at 1931 Å and 2478 Å, CII (*i.e.*, C$^+$ in spectroscopic notation) at 2837 Å and a host of other emission lines some of which are indicated. Ne$^+$ lines are shown.

# 5 Mechanisms and parameters for the formation of the regenerative soot

## 5.1 Potential and kinetic sputtering of the sooted cathode

We compare the contributions of the two mechanisms of emission of C and its clusters by kinetic and potential sputtering. The relative contributions of the two wall removal processes is essential to understand the state of the carbon vapor in the regenerative sooting discharge. Both play their respective roles in the initial sputtering of the cathode and the later regeneration of the soot. Kinetic sputtering generates linear collision cascades by the incident ions which on interaction with the surface can emit the surface constituents as sputtered particles into the plasma. We use SRIM2000 [19] for obtaining the kinetic sputtering yield $k = 0.12 \pm 0.02$ C$_1$ atoms/Ne$^+$ with 500 eV energy incident on graphite. The potential sputtering of the sooted surface can take place upon the interaction of the Ne$^*$ metastable atoms with 16.7 eV. Similarly, in the later stages C$^*$ or C$^*$ also are likely to contribute effectively to the potential sputtering. Either an individual C$_1$ or a whole cluster C$_m$ ($m \geq 2$) of many atoms adsorbed on the sooted cathode with binding energies $E_{C_m} < E_{(Ne^*; C_1^+; C_2^+)}$, can be ejected with the interaction of the excited species. The clusters that are recycled can further go through the process of disintegration into smaller units or fragments. In our experiments the experimentally observed ratio is CI/Ne$^+$ = 4.5 ± 30% in the pressure range 0.1−1 mbar and is relatively insensitive to the variations of the discharge current $i_{dis}$. The ratio of the excited C and Ne is CI/NeI × 10$^{-3}$. Using the kinetic sputtering yield data of SRIM2000 for the ratio C1/NeII, we get $k \sim 0.12$ C$_1$/Ne$^+$. Thus the kinetic sputtering will provide ~10$^{-5}$ CI excited atoms in the $^1P_1$ level. This is much smaller than the observed relative density CI ≈ 4.5 × Ne$^+$. The potential sputtering on the other hand, can also release between 0.5 and 1 CI per NeI. This would yield ~ 10$^{-2}$ excited CI atoms for each NeI metastable atom, as the sputtering is occurring from a sooted cathode that is covered with many monolayers of loose agglomeration of clusters. These clusters contain on the average 50−100 C1 atoms [1]. The potential sputtering of the clusters estimates are within 50% of the observed value. The high number density of CI that is observed is primarily due to the potential sputtering of the carbon clusters adsorbed on the sooted surface by the metastable atoms and clusters. Therefore, potential sputtering is the dominant soot regeneration mechanism in graphite hollow cathode carbonaceous discharges that are started and sustained with noble gases.

## 5.2 Role of the two energy regimes of electrons in hollow cathode glow discharges

|  | 1 eV | 10 eV | 100 eV |
|---|---|---|---|
| CI–CII | $1.93 \times 10^{-13}$ | $1.7 \times 10^{-8}$ | $9.7 \times 10^{-8}$ |
| CII–CIII | $1.2 \times 10^{-19}$ | $1.3 \times 10^{-9}$ | $2.6 \times 10^{-8}$ |
| CIII–CIV | $4 \times 10^{-30}$ | $5.8 \times 10^{-11}$ | $7.4 \times 10^{-9}$ |
| NeI–NeII | $1.15 \times 10^{-18}$ | $1.32 \times 10^{-9}$ | $4.2 \times 10^{-9}$ |
| NeII–NeIII | $3.25 \times 10^{-28}$ | $9 \times 10^{-11}$ | $1.8 \times 10^{-8}$ |

**Table 1.** Ionisation rate coeffcient $\alpha_i$ cm$^3$ s$^{-1}$ for the atomic and ionic species of C and Ne, CI-CII, CII-CIII, CIII-CIV and NeI-NeII, NeII-NeIII. All ionisations are from the ground state.

The levels of excitation and ionization of the support gases and the sputtered species cannot be explained by a single electron energy regime *i.e.*, the excitation temperature $T_{exc}$ of any two levels of a species. It is well documented that the hollow cathode glow discharges are initiated and sustained by two well defined electron energy regimes [20]. In our case the higher regime has $E_e \geq 10$ eV while the other electron energy range can be evaluated from the excitation temperature obtained from the emission lines of NeI, NeII and, if possible, from CI and the excited C$_2$. These provide us an average kinetic energy of electrons $\langle E_e \rangle \approx T_{exc} \leq 1\ eV$. The role played by these two distinct energy regimes of electrons can provide an explanation of the rather high densities of the ionized species CII, CIII, NeII and NeIII that are present even at $i_{dis}$ as low as 50 mA. Table 1 is prepared by using Lotz's' semi empirical formulation [21] for the ionization rate coefficients $\alpha_i$ cm$^3$ s$^{-1}$ for the successively higher ionization stages of C and Ne. For these calculations Maxwellian velocity distribution for the electrons is assumed and all excitations and ionizations are from the ground state. At $T_e \approx 1$ eV or less which can be approximated as the discharge temperature in our case, the presence of the higher ionized species is much less probable. However, at higher electron energies $E_e \geq 10$ eV, a significant increase in $\alpha_i$ occurs. Between $10-100$ eV energy range, the electrons can ionize all ionic stages of C and Ne with similar orders of magnitude probabilities. The spectral line ratios of the C ions have been used to evaluate $T_e$ for the carbon impurities in the tokomak plasmas for CII to CIV in the $T_e \sim 4 - 40$ eV range. Since in our case the $T_e \sim 1$ eV, the required high energy electrons are provided by the cathode for the ionization for the higher ionization levels of C and Ne. These are available due to the secondary electron emission from the cathode but in much reduced intensities compared with the thermal electrons.

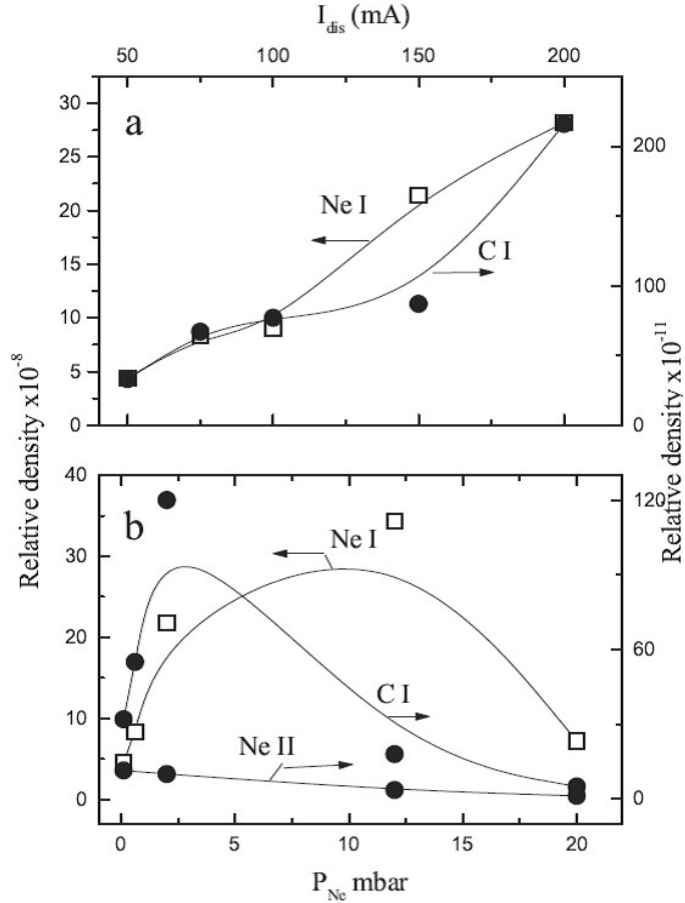

**Fig. 5.** The relative densities of NeI, CI, NeII are shown as a function of the discharge current $i_{dis}$ in (a) and as a function of $P_{Ne}$ in the range 0.6 to 20 mbar in (b).

In Figure 5 we have plotted the relative number densities of NeI, NeII and CI as a function of the discharge current $i_{dis}$ in Figure 5a and as a function of the gas pressure $P_{gas}$ in Figure 5b. The average relative number densities of NeII/NeI in Figure 5a is $\sim 1.3 \times \pm 0.3 \times 10^{-3}$. The carbonaceous character is determined by the ratio CI/NeI ~ $0.6\pm0.1\times10^{-2}$. The most conspicuous feature of all photoemission spectra is the presence of strong CII intercombination multiplet at $\lambda$ = 2324-2328 Å. We will discuss its possible origin in the next section. However, the ratio CII/CI ~ $3.5\pm0.6 \times10^{4}$! It identifies that higher energy electrons have to be present to ionize and further excite these ions. Table 1 can highlight the significance of the higher energy electrons.

## 5.3 The state of atomic and ionised C

Figure 6 shows two spectra with the source gas pressure $P_{Ne} \approx 0.6$ mbar. During the first spectrum in Figure 6a with $i_{dis}$ = 200 mA, the discharge voltage $V_{dis}$ = 0.6 kV. This spectrum is taken on a freshly prepared cathode surface, at high values of $V_{dis}$ and $i_{dis}$. This is a typical carbon cathode sputtering dominated spectrum. All the respective neon lines both atomic and ionic are present. But the VUV part of the spectrum is dominated by carbon's neutral, singly and doubly charged CI, CII and CIII lines. In Figure 6b the emission lines $\lambda$ = 1931 Å and $\lambda$ = 2478 Å are the two dominant VUV lines of CI that can be seen along with the $\lambda$ = 2 135 Å and $\lambda$ = 2191 Å of CII and the $\lambda$ = 2297 Å of CIII. Also present is CII's 233 nm intercombination multiplet. It has very small transition probabilities ~0.1 s$^{-1}$ for all five lines of the multiplet. This multiplet is generally a weak intersystem transition route for the de-excitation of CII in carbon plasmas. From the energy level scheme of C in NIST Database [18], out of the total of 254 CII emission lines between 0−2000 Å, 73 lines are emitted by de-excitation to the first excited level $2s2p^2$ $^4P_{(1/2;3/2;5/2)}$ of CII. This level, in turn de-excites to the ground level $^2P_{(1/2;3/2)}$ by the emission of the 232 nm multiplet. The intense emission indicates that CII exists as a highly excited C ion in the discharge. The NeI lines between 580−650 nm remain as the significant emission feature of all of these spectra.

Singly ionized carbon's first excited state $^4P_{(1/2;3/2;5/2)}$ has lifetime $\tau \approx 4.7$ ms. Thus its level density serves as a useful indicator of the carbon content of the cusp field, graphite hollow cathode plasma. The calculated relative densities from the line intensities of CI $\lambda$ = 1931 Å, CII $\lambda$ = 2324−2328 Å, NeI $\lambda$ = 5852 Å, NeII $\lambda$ = 3713Å provide us the statistics of the carbonaceous discharges. From the natural radiative lifetimes of these four excited states CII has six orders of magnitude longer residence time in the plasma. We already have discussed earlier that its collisions with the walls are most likely as opposed to the short lived constituents. In the $i_{dis}$ range 50−200 mA we get the ratio of the densities $D_{CII}/D_{CI}$ ~$(3.5\pm0.5)\times10^4$. Similarly, $D_{NeII}/D_{NeI}$ ~ $(1.3\pm0.3)\times10^{-3}$ and $D_{CI}/D_{NeI}$ ~ $(0.6\pm0.1)\times10^{-2}$. These results identify a carbonaceous plasma with the ionised C whose ratio with the excited Ne is $D_{CII}/D_{NeI}$ ~ $2\times10^2$. The results also imply that the density of the singly ionised neon $D_{NeII}$ is only ~$10^{-5} \times D_{CII}$ throughout our discharge current range. Therefore, our carbonaceous discharge is dominated by the ionised C and has a 2−4% excited Ne.

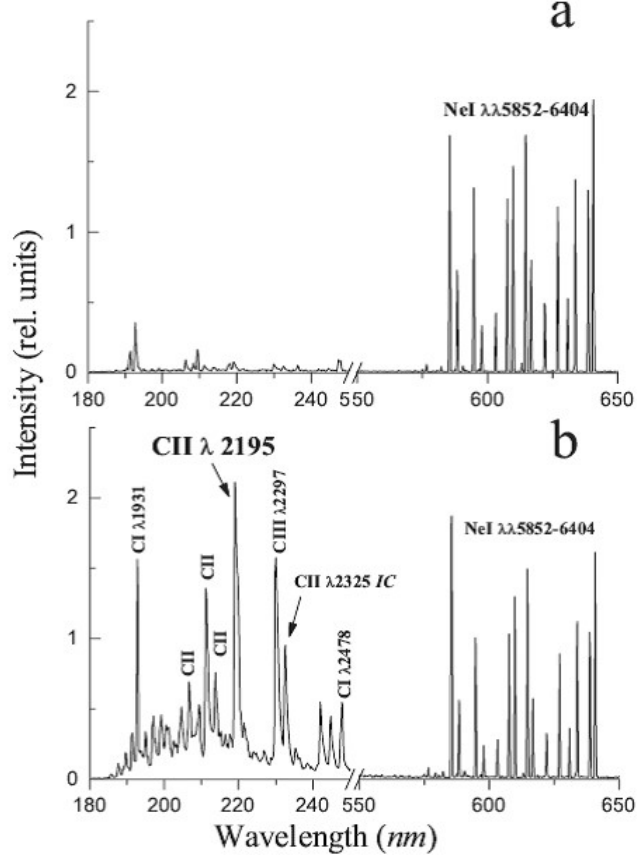

**Fig. 6.** Photon emission spectra are presented with Ne as the support gas at two different discharge voltages with $P_g$ ≈0.6 mbar, $i_{dis}$ = 200 mA being kept constant. For (a) $V_{dis}$ = 0.6 kV and (b) $V_{dis}$ = 1.5 kV. The relative lines intensities are plotted against the wavelength. The x-axis is broken in the two wavelength ranges of 180−250 nm and 50−650 nm, respectively.

## 5.4 $T_{exc}$ in the carbonaceous, non-LTE discharge

A multi-component glow discharge whose composition is in a state of flux in general, cannot be in local thermodynamic equilibrium-LTE. The ongoing processes of the soot regeneration that involve formative as well as fragmentation stages of clusters continuously modify the discharge composition. From such non-LTE plasma, we can only calculate the excitation temperatures of different species rather than the typical electron temperature $T_e$. The ratios of the upper ($N_u$) and lower ($N_l$) level densities for respective transitions to the same level $m$. The line intensities $I_{um} = N_u A_{um} h\nu_{um}$ and a similar one for $I_{lm}$. The two intensities involved three levels are used to for the evaluation of the relative level densities $N_u=N_l$. The third level is the common one for the transitions to the same lower level $m$. The ratio of the two level densities is given by [22] as $N_u=N_l = (g_u=g_l) \exp\{-(E_u - E_l)=kT_{exc}\}$, where $g_u$, $g_l$ are the statistical weights, $E_u$ and $E_l$ the energies of the respective levels. An energy difference ~1−2 eV between the two upper terms yields higher accuracy in the determination of $T_{exc}$.

The excitations in the carbonaceous plasma's atomic and ionic species are induced by electron collisions. Assuming Maxwellian velocity distribution for the electrons, their temperature $T_{exc}$ can be evaluated from the two resonant lines of NeI $\lambda$ = 5852 Å and $\lambda$ = 3520 Å. The level $3s[1/2]$ at 16.85 eV is populated by the spontaneous emission of these two lines from $3p[1/2]$ and $4p[1/2]$ at 18.96 and 20.37 eV, respectively. This provides $T_e \approx 8500 \pm 300$ K for the discharge current $i_{dis}$ in the range 50−200 mA. While by using two non-resonant transitions NeI $\lambda$ = 3501 Å and NeI $\lambda$ = 5400 Å we get $T_e \approx 5700 \pm 500$ K for the same $i_{dis}$ range. The excitation temperature for the two sets of the singly charged neon lines NeII is a factor of 2 higher than that obtained by using NeI lines. With NeII $\lambda$ = 1938 Å and NeII $\lambda$ = 3727 Å we get $T_e \approx 15800 \pm 1000$ K. All sets of these transitions are, respectively, to the same lower levels. $T_{exc}$ are also obtained for those discharge conditions where the cathode surface is freshly prepared *i.e.*, no prior sputtering takes place and one gets the emissions from an un-sooted surface as a function of the pressure.

For the sake of comparison, the resonant transitions of NeI $\lambda$ = 5852 Å and $\lambda$ = 3520 Å yield three values of $T_{exc} \approx$ 16300 K ($P_{Ne}$ = 0.06 mbar), 10700 K ($P_{Ne}$ = 0.1 mbar), and 8500 K ($P_{Ne}$ = 2 mbar), respectively. Our data indicates a remarkable consistency in the profile and relative intensity ratio of the resonant emission lines NeI $\lambda$ = 5852 Å and $\lambda$ = 3520 Å for a well-sooted discharge. The ratio of the densities of the excited carbon to neon $D_{CI}/D_{NeI} \approx (0.6 \pm 0.15) \times 10^{-2}$. The measured ratio of the singly ionized to the neutral neon $D_{NeII}/D_{NeI} \approx (1.3 \pm 0.3) \times 10^{-3}$. The relative intensity of the CII 232 nm intercombination multiplet and its ratio to NeI and NeII is as $D_{CII}/D_{NeI} \sim 10^2$ and $D_{CII} \sim 10^5 \times D_{NeII}$. The most dominant ionized species in the discharge is therefore, CII.

# 6 The C$_2$ and C$_3$ content of the regenerative soot

C$_2$ and C$_3$ are the essential ingredient of the sputtered species from graphite as well as from the sooted discharges. But the relative ratios of their neutral and charged states show large variations. These variations depend on the underlying mechanisms that are operating. Honig [3] had mass analyzed the subliming clusters from a graphite oven and measured the ratio of the relative densities of the positively charged atoms and clusters as C$_1$:C$_2$:C$_3$ = 1:0.37:2.83, While for the negative species $C_1^-:C_2^-:C_3^-$ = 1:3600:40. In our experimental results shown in Figure 3 we saw that velocity spectra of the positively charged carbon clusters $C_m^+$ ($m \geq 2$) from the regenerative sooting discharge is dominated by clusters with large carbon content *i.e.* $C_m^+$ ($m \geq 4$). $C_2^+$ is present but only as a minor constituent. On the other hand, the positively charged clusters $C_m^+$ that are kinetically sputtered from a flat graphite disc are likely to auto-neutralise. That may be the reason that we could detect only negative clusters in a non-regenerative environment. We could detect $C_2^-$ along with $C_3^-$ and $C_4^-$ in the mass spectrum of the sputtered graphite species [2].

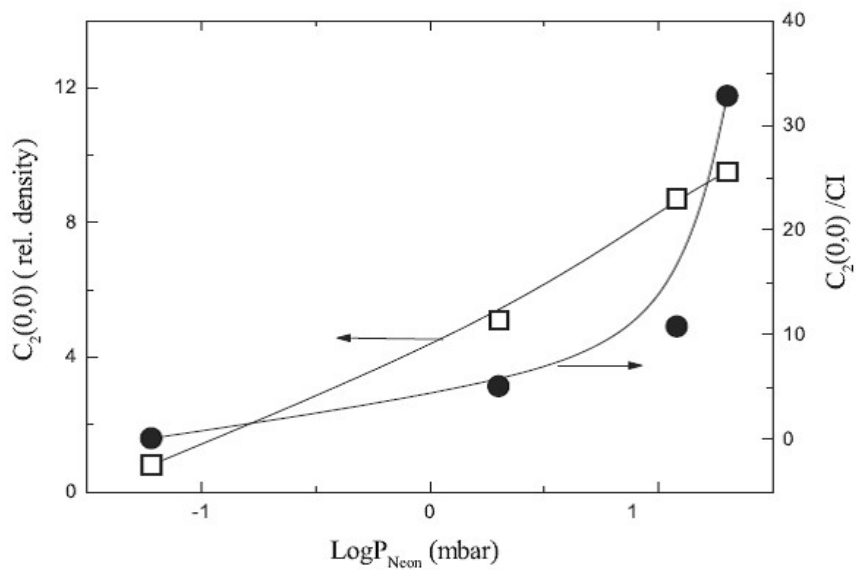

**Fig. 7.** The emission spectra yield the number densities of the vibrationally excited $C_2(0; 0)$ at $\lambda$ = 5165 Å and its ratio with the excited C atoms CI at $\lambda$ = 2478 Å is plotted as a function of $P_{Ne}$ between 0.06 and 20 mbar at $i_{dis}$ = 75 mA. The $C_2(0; 0)$/CI ratio rises steeply after $P_{Ne}$ > 1 mbar.

An emission spectrum from the regenerative sooting discharges was shown in Figure 4. In similar experiments the Swan band heads of $C_2$: $C_2$ (0, 0) at $\lambda$ = 5165 Å, $C_2$ (1, 0) at $\lambda$ = 4737 Å, $C_2$ (2, 1) at $\lambda$ = 4715 Å, and $C_2$ (0, 1) at $\lambda$ = 6535 Å have been observed. Therefore, the Swan band of $C_2$ is a useful indicator of $C_2$ in the sooting discharges. An increase in the number density of neutral $C_2$ as a function of the discharge current $i_{dis}$ follows a similar pattern of direct proportionality to $i_{dis}$ by the atomic species $C_1$. Whereas, on increasing the support gas pressure from 0.06 to 20 mbar at constant $i_{dis}$ as shown in Figure 7, we notice an inverse relationship between the number densities of the atomic and diatomic carbon species. We have derived the number densities of the excited $C_1^*$ and $C_2^*$ from the line intensities of the electronically excited CI at $\lambda$ = 2478 Å and the lowest transition of the Swan band with the vibrationally excited $C_2(0, 0)$ at $\lambda$ = 5165 Å. $C_2(0, 0)$ number densities are plotted on the left vertical axis and the ratio $C_2(0, 0)$/CI along the right vertical axis. $C_2(0, 0)$ is an increasing function of both of the discharge parameters $i_{dis}$ and $P_{Ne}$. CI also increases rapidly with $i_{dis}$, but its contribution is reduced at higher pressures as can be seen in Figure 5. From the photon emission data the ratio of the singly charged to the excited C $i.e.$ CII/CI remain constant, for example in the range of $P_{Ne}$ = 0.1−1 mbar it is 0.55±0.2 for $i_{dis}$ = 50 to 200 mA. Similarly $C_2(0, 0)$/CI = 2.2 ± 0.4 under the same conditions. At low pressure discharge $i.e.$, $P_{Ne} \approx$ 1 mbar the charged atomic carbon $C_1^*$ (i.e., CII) and the excited diatomic molecular carbon ($C_2(0, 0)$) are directly related with CI $i.e.$, $C_1$ in the $^1P_1$ level ($E^1P_1 \approx$ 7.5 eV). Thus it may be deduced that the origin of $C_1$ and $C_2$ is in the dissociation of $C_3$ $via$ $C_3 \rightarrow C_2 + C_1$.

# 7 The transition from the C₃ dominated discharge to the sooting plasma

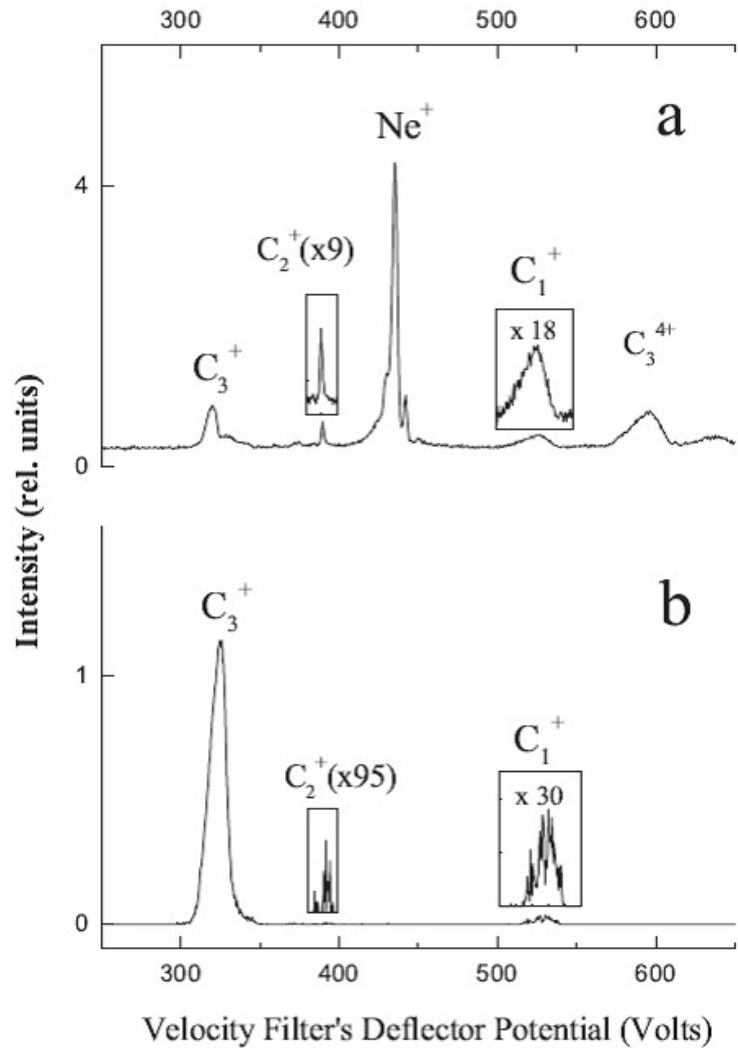

**Fig. 8.** (a) $i_{dis}$ = 150 mA, Ne⁺ is the most significant peak followed by $C_3^+$ and $C_3^{4+}$. Inset: $C_2^+$ and $C_1^+$ are shown magnified by a factor of 9 and 18, respectively. (b) $i_{dis}$ = 12.5 mA, $C_3^+$ is the only and most significant peak. Inset: $C_2^+$ and $C_1^+$ are shown by their respective enlargement factors.

The mass spectrum for the positively charged clusters emitted from the source operated at low pressures and a well confined discharge has certain unique characteristics. It is dominated by $C_3^+$ and much smaller densities of $C_1^+$ and $C_2^+$. Two such spectra are shown in Figure 8 by confining the discharge within the annular region between the hollow cathode and the extended hollow anode (see Fig. 1a). The physical confinement between the thin annular region is supplemented by the cusp magnetic field $B_z(r, \theta)$ contours (see Fig. 1c). The two spectra shown in Figure 8 are with $i_{dis}$ = 150 mA (Fig. 8a) and $i_{dis}$ = 12.5 mA (Fig. 8b). We notice that $C_3^+$ is the major surviving species and at low $i_{dis}$, it is discharge's main ionic component. However, in the photoemission spectra from at these conditions *i.e.* $i_{dis}$ = 12.5 mA, we have seen the atomic and ionic C lines (CI, CII, C₂ etc.). The velocity spectrum at $i_{dis}$ = 150 mA shows $C_3^+$ and Ne⁺ as the main discharge features. Reducing $i_{dis}$ by a factor of 20 to $i_{dis}$ = 12:5 mA, the $C_3^+$ is the only surviving and ionized species. In Figure 8 the peaks due to $C_2^+$ are enlarged by factors of 9 and 95, respectively. Whereas, the $C_1^+$ intensity is seen to be enlarged by 18 and 30 times in the two respective figures. Broad peak due to $C_3^{4+}$ is also present in the spectra.

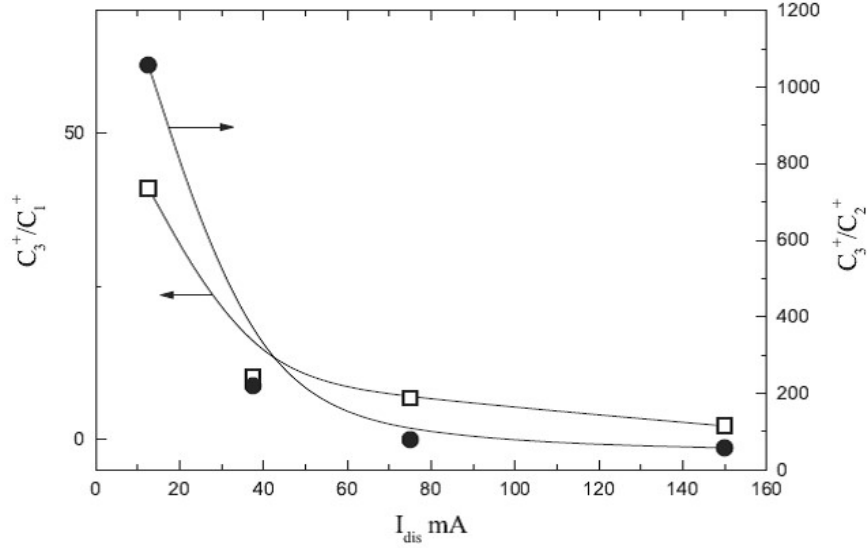

**Fig. 9.** The ratios of the ion densities $C_3^+/C_1^+$ and $C_3^+/C_2^+$ are plotted as a function of $i_{dis}$. $C_3^+$ is the main discharge constituent and its contribution increases by lowering $i_{dis}$.

The accumulated data for the relative ion densities of $C_3^+$, $C_2^+$ and $C_1^+$ as functions of $i_{dis}$ and $P_{Ne}$ are presented in Figure 9. The ratios of $C_3^+/C_1^+$ and $C_3^+/C_2^+$ are plotted as a function of $i_{dis}$ from 12.5 to 150 mA. $C_3^+$ is the sole survivor at very low discharge currents. Its relative number density increases with respect to that of $C_1^+$ by a factor of 27 for decreasing $i_{dis}$ = 150 mA to 12.5 mA. Whereas, $C_3^+/C_2^+$ increases by two orders of magnitude as $i_{dis}$ is decreased in the same range. The number density of $C_2^+$ increases from 5% to 20% of $C_1^+$. From the tabulated data presented in Figure 9 we conclude that $C_3^+$ is not only the significant species at low $i_{dis}$ but is the main constituent of the discharge under the experimental conditions. We have recently argued [23] that the predominance of $C_3$ in such a discharge may be due to the regeneration pattern of all clusters $C_m^+$ ($m > 3$) up to $C_{30}$ favors the accumulation of $C_3$ as the end product. This fragmentation scheme $C_m^+ \rightarrow C_{m-3}^+ + C_3$ with dissociation energy $E_{dissoc} \approx 5.5\pm0.5$ eV has been predicted and experimentally verified for all $C_m^+$ up to $m \leq 60$ [24-26]. Their end product $C_3$ can itself dissociate into $C_2$ and $C_1$ via $C_3 \rightarrow C_2 + C_1$ with $E_{dissoc} \approx 7\pm0.5$ eV. We invoke this fragmentation pattern to explain the preponderance of $C_3$ as well as the enhanced contribution of $C_1$ at higher $i_{dis}$. At higher source pressures $P_{Ne}$, the large relative increase in the density of $C_2$ cannot be explained by the $C_3$ fragmentation alone. For example at $P_{Ne} > 1$ mbar, the increased contribution from the $C_2(0, 0)$ is accompanied by a consequent decrease in the excited and ionic C1 (CI, CII, CIII) lines. We interpret it as the onset of the formation of the closed caged clusters $C_m$ ($m > 30$). These clusters further fragment *via* C$_2$ emission $C_m \rightarrow C_{m-2} + C_2$ ($m \geq 30$). This is why we observe enhanced C$_2$(0, 0) intensities at $P_{Ne} > 10$ mbar.

## 8 Conclusions

In this study, we have shown that the simultaneous mass spectrometry of the clusters extracted from the source and the photon emission spectroscopy of the carbonaceous discharge yields important information about the formation, dissociation and fragmentation of clusters within the sooting discharge. All these sequences constitute the essential stages of the soot formation. The main agents for the regeneration of the soot are;
(i) the kinetic and potential sputtering from the sooted cathode,
(ii) collisions with energetic electron,
(iii) collisions between the metastable $Ne^-$, $C_1^-$, $C_2^-$, $C_3^-$, ...constituents of the discharge.

The inclusion of carbon into the neon plasma has been clearly identified from the data on the relative number densities of CI and NeI calculated from their characteristic lines' intensities in the emission spectra. The initiation of the discharge requires high $T_{exc}$ regimes with higher values of $V_{dis}$ and $i_{dis}$. The sputtering of the graphite cathode with the subsequent excitation and ionisation of the carbon atoms with high energy electrons emitted from the cathode and accelerated in the cathode fall to energies ~ 500 eV. It is the main agent for the discharge sustenance with gradually increasing C content. The singly ionised CII content participates efficiently in the kinetic sputtering of the cathode along with NeII. Due to the presence of the large fraction of the negative species, the possibility of $C_x^+ + C_y^- \rightarrow C_{x+y}^{0,\pm}$ type interactions may be a significant step towards larger cluster formation.

We have shown that the soot formation in the hollow cathode discharge may proceed in two distinct stages:
(1) sputtering dominated regime with the discharge produced and contained in the annular region between the cylindrical anode and cathode. This non-LTE discharge is dominated by C$_3$ which is the sole survivor of the linear chains and ring type clusters $C_m$ ($m \leq 30$). C$_3$ itself is the end-product of the $C_m$ fragmentation in the discharge. Further dissociation of C$_3$ into C$_1$ and C$_2$ provides a highly excited and ionized C$_1$ as indicated by the emission lines of the neutral, singly and doubly charged (CI, CII and CIII) species from such a discharge;
(2) if the discharge is not constrained and allowed to expand into the extended hollow cathode, one obtains the soot formative environment. Larger clusters $C_m$ ($m \geq 30$) are formed and we obtain the cage closure resulting in the formation of fullerenes. The cage closure amounts to carbon accretion at a very rapid rate during the sooting stages as opposed to the sputtering one described above.